\documentclass[conference]{IEEEtran}
\IEEEoverridecommandlockouts
\usepackage{natbib}
\usepackage{amsmath,amssymb,amsfonts}
\usepackage{algorithmic}
\usepackage{graphicx}
\usepackage{textcomp}
\usepackage{xcolor}
\def\BibTeX{{\rm B\kern-.05em{\sc i\kern-.025em b}\kern-.08em
    T\kern-.1667em\lower.7ex\hbox{E}\kern-.125emX}}

\begin{document}
\title{Comparative Analysis of CPU and GPU Profiling for Deep Learning Models
}

\author{\IEEEauthorblockN{Dipesh Gyawali}
\IEEEauthorblockA{\textit{Department of Computer Science} \\
\textit{Louisiana State University}\\
Baton Rouge, Louisiana, United States \\
dgyawa1@lsu.edu}
}

\maketitle
% 2nd. author
%\alignauthor
%G.K.M. Tobin\titlenote{The secretary disavows
%any knowledge of this author's actions.}\\
%       \affaddr{Institute for Clarity in Documentation}\\
%       \affaddr{P.O. Box 1212}\\
%       \affaddr{Dublin, Ohio 43017-6221}\\
%       \email{webmaster@marysville-ohio.com}

% There's nothing stopping you putting the seventh, eighth, etc.
% author on the opening page (as the 'third row') but we ask,
% for aesthetic reasons that you place these 'additional authors'
% in the \additional authors block, viz.
%\additionalauthors{Additional authors: John Smith (The Th{\o}rv{\"a}ld Group,
%email: {\texttt{jsmith@affiliation.org}}) and Julius P.~Kumquat
%(The Kumquat Consortium, email: {\texttt{jpkumquat@consortium.net}}).}
% Just remember to make sure that the TOTAL number of authors
% is the number that will appear on the first page PLUS the
% number that will appear in the \additionalauthors section.

\maketitle
\begin{abstract}
Deep Learning(DL) and Machine Learning(ML) applications are rapidly increasing in recent days. Massive amounts of data are being generated over the internet which can derive meaningful results by the use of ML and DL algorithms. Hardware resources and open-source libraries have made it easy to implement these algorithms.  Tensorflow and Pytorch are one of the leading frameworks for implementing ML projects. By using those frameworks, we can trace the operations executed on both GPU and CPU to analyze the resource allocations and consumption. This paper presents the time and memory allocation of CPU and GPU while training deep neural networks using Pytorch.  This paper analysis shows that GPU has a lower running time as compared to CPU for deep neural networks. For a simpler network, there are not many significant improvements in GPU over the CPU.
\end{abstract}

%A category including the fourth, optional field follows...
%\category{D.2.8}{Software Engineering}{Metrics}[complexity measures, performance measures]


%\keywords{ACM proceedings, \LaTeX, text tagging}

\section{Introduction}
Machine Learning is a field that makes the computer system learn without being explicitly programmed. It requires a massive amount of data to generate meaningful information and make decisions. The formal and scientific definition of Machine Learning is "A computer program is said to learn from experience E with respect to some class of task T and performance P. If its performance at tasks in T, as measured by P, improves with experience E."\cite{ml} Deep Learning is a sub-branch of machine learning which tends to use neural networks for data processing and decision making. The working of artificial neural network models mimics the human brain. Deep neural networks need a lot of computations and hardware resources for proper operations as huge amounts of matrix and algebraic operations take place. 

Central Processing Unit (CPU) and Graphical Processing Unit (GPU) are two processing units that are extensively used to process ML and DL models. GPU is specially designed for parallel computation while CPU is not used for the same. Pytorch\cite{pytorch} and Tensorflow\cite{tensorflow} are two frameworks that provide an abstraction for complex mathematical calculations in the field of deep learning. 

The need for GPUs arises for training the deep learning network when a vast amount of data is generated. The normal CPUs typically have 4-5 cores and only a limited number of threads can be handled. While GPUs have thousands number of small cores that can handle several computation threads in parallel as shown in Figure 1. For example, NVIDIA A6000 has 10752 CUDA cores. A deep learning system contains millions of calculations to train and infer which needs GPU for fast processing.

\begin{figure}[hbt!]
  \centering
  \includegraphics[scale=0.25]{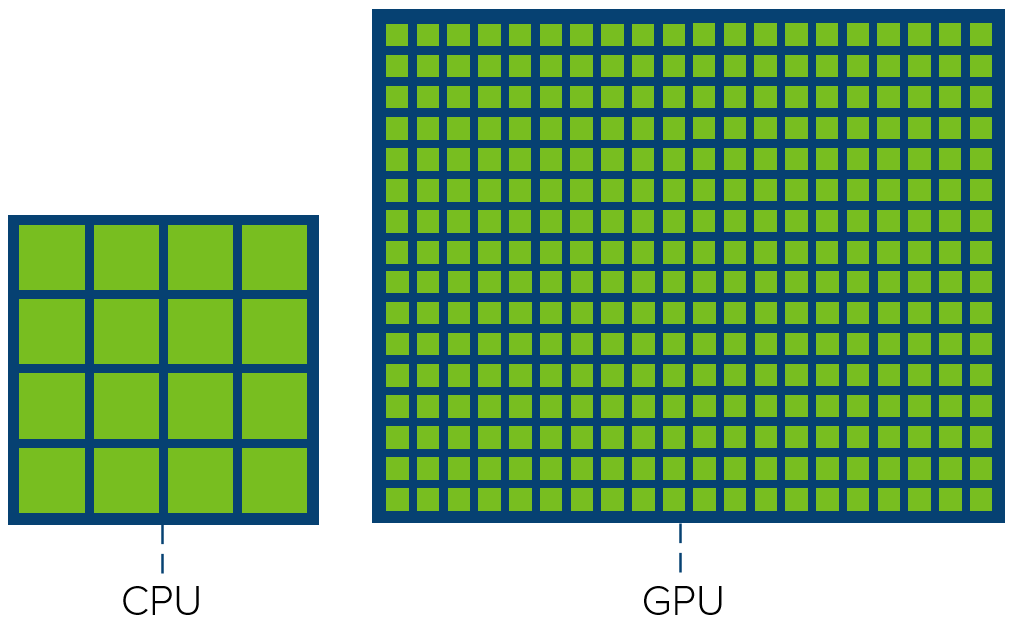}
  \caption{CPU and GPU}
  \end{figure}

\section{Problem Statement}
In this research, we will focus on how the mathematical operations are executed on CPU and GPU and analyze their time and memory. Analyzing time and memory at runtime helps to optimize the network operations which helps in faster execution and inference. Using Pytorch Profiler\cite{pytorch} API and NVIDIA commands, we can visualize the network graphs and find out the execution time of each operation in addition to the memory consumption status

\section{Scope}
\vspace{0.5cm}
The main motivation behind this research is to find how CPU and GPU operations happen while training machine learning and deep learning models. The main focus will be on CPU and GPU time and memory profiling part, but not on the deep learning models. The profiling will be generated using a deep learning model using Pytorch\cite{pytorch} profiler and Tensorboard\cite{tensorflow}. Tracing is done at each training step to get the data about time and memory. Later, the resource utilization of CPU and GPU is compared. 

\section{Literature Review}
A number of works have been done to identify CPU and GPU performances over different algorithms and operations. Lind et al.\cite{lind} talk about the performance comparison between CPU and GPU in Tensorflow and Gyawali et al.\cite{gyawali2020comparative} talked about building deep CNN models across GPU for different algorithms. They have used the Tensorflow profiler to trace the operations in CPU and GPU. Similarly, Coutinho et al.\cite{coutinho} mention about the profiling strategy based on performance predicates and identify major causes of performance degradation. In the same way, Alkaabwi\cite{intisar} uses Tensorflow and Big Dataset to compare CPU and GPU neural network parallel implementation. Salgado\cite{ronie} describes profiling kernel behavior to improve CPU/GPU performances. Satish et al.\cite{satish} mention the ways to optimize the GPU memory for large-scale datasets. Likewise, Stephenson et al.\cite{stephenson} describe the flexible software profiling of GPU architectures.

\section{Background}
\subsection{Terminology}
The following terminology is used in the paper.
\begin{itemize}
\item Neural Network: It is the group of interconnected nodes where the data flow and processing happens to produce an output.
\item CPU: CPU stands for Central Processing Unit which processes the task inside a computer. It consists of RAM, a cache, a control unit, and an arithmetic unit.
\item GPU: GPU stands for Graphical Processing Unit that consists of multiple components similar of CPU designed for parallel computations.
\item CUDA: CUDA stands for Computer Unified Device Architecture which is a parallel computing platform and model used to make programs compatible to work with GPU.
\item Profiler: It is a tool to analyze and measure the performance of the program, and usage of resources like time and memory.
\item Memory Allocation: The process of allocating computer memory to execute programs and processes.
\item Pytorch: It is a deep learning framework developed by Meta which helps us to build neural network pipelines.
\item Tensorflow: It is also a framework to work on deep learning applications created by Google.
\end{itemize}

\subsection{CPU}
The primary responsibility of the CPU is to execute the task in the computer memory sequentially. It has different components i.e. arithmetic unit, and logical unit. During neural network computations, the CPU also plays a major role in calculating general arithmetic calculations. It doesn't have a high number of cores as compared to a GPU. The cores are clocked at an average of 2 to 3 GHz which makes the CPU ideal for processing sequential tasks. The data loader uses the CPU to load the training data that is handled by the CPU. The I/O operations are also handled by the CPU, which makes it an important component in a computer system. 

\subsection{GPU}
GPU is similar to CPU with the difference that it can compute a larger amount of work in parallel at the same time. GPU is one of the most important components to train a deep neural network model. It has thousands of cores that help to distribute the task and work in parallel. It has a low clock rate as compared to the CPU. It uses thread to process higher-dimension calculations, which can speed up the computations. Several GPUs are available in the market. In this experiment, we used 2 NVIDIA A6000 GPUs named GPU:0 and GPU:1, the specification of which is shown in Figure 2.

\begin{figure}[hbt!]
  \centering
  \includegraphics[scale=0.20]{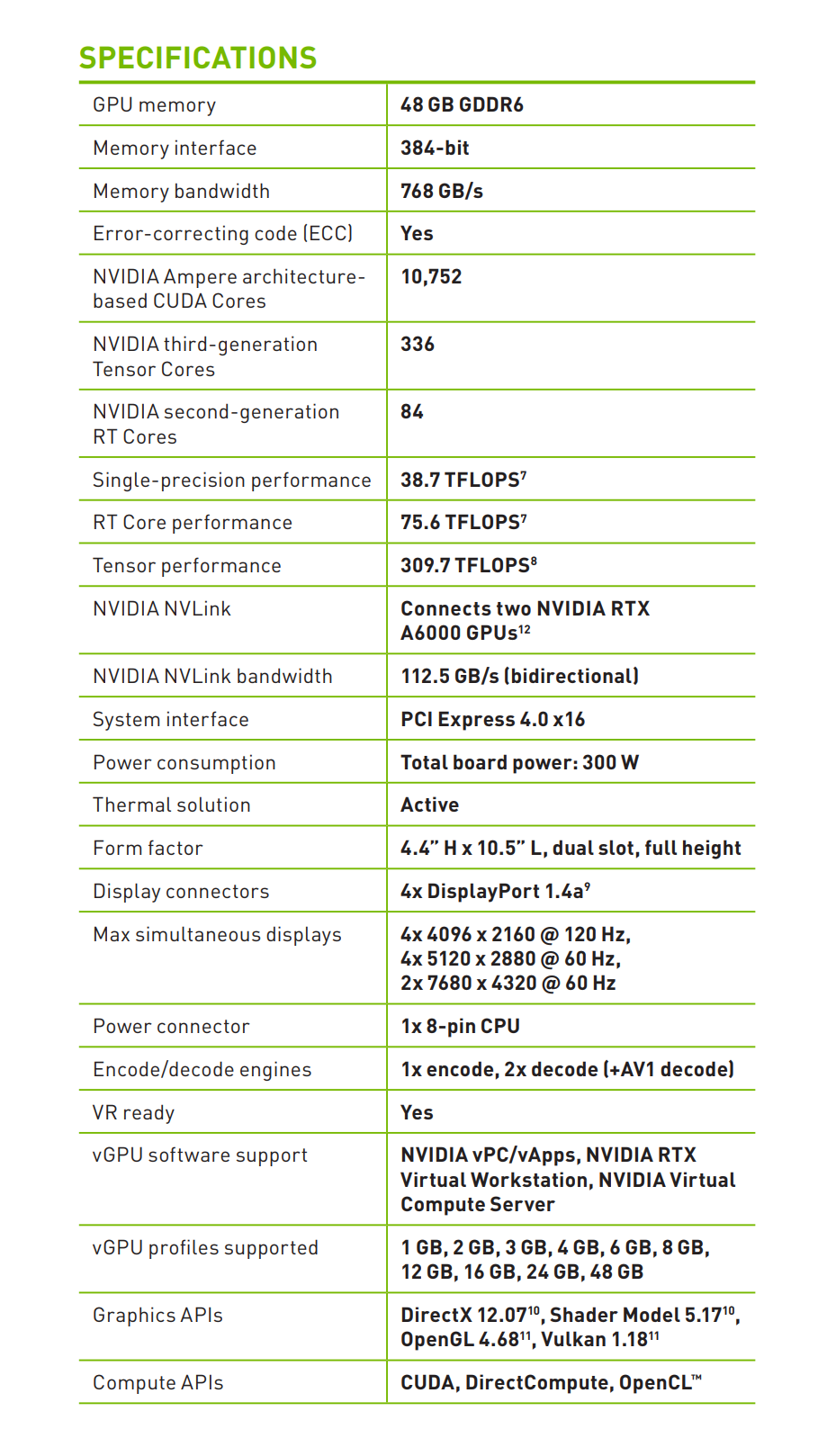}
  \caption{NVIDIA A6000 specification}
  \end{figure}

Basically, GPUs are initially built for rendering and visualizing graphical images and videos. It requires API to use GPU operations at a higher abstraction level. Its architecture offers various parallelisms like multi-threading, SIMD, and MIMD. So, Cuda is designed to use low-level instructions for utilizing these kinds of parallelism. Pytorch\cite{pytorch} is compatible with CUDA to use GPU architecture to profile and optimize the operations.

\subsection{Profiling}
Software profiling is done to achieve the best performance of a system. Machine learning and deep learning applications take a lot of time to train and consume a lot of memory. Profiling is a way to find how the code operates in response to time and memory. Profiling is extremely important to optimize any deep neural networks so that we can achieve faster way computations. It is important to understand CPU, GPU, and memory bottlenecks, which can cause slowdowns in training and inference of models.

Resource monitoring helps us identify how much memory is consuming to play with different metrics to produce optimal results. When using GPU, we want almost every operation to be handled by GPU when training and inferring using a deep neural network model. By using the profiling techniques, we can find out how much GPU is utilized by the system to make the code run on GPU for higher efficacy. Pytorch Profiler\cite{pytorch} can help us to identify how much time and memory is being used by CPU and GPU. In addition to this, we also used various commands from NVIDIA to evaluate the profiling metrics.

\section{Metrics}
GPU monitoring should be done constantly while training and deploying deep learning applications. The following metrics are mostly monitored for deep learning networks:

\subsection{GPU Utilization}
When training a deep learning model, GPU utilization is a metric that needs constant observation if the model is utilizing the GPU. It refers to the percentage of time one or more GPU kernels are running over the last second, which is the same as GPU being utilized by a deep learning program. It helps us to determine whether GPU is being used fully by the operations. In addition to this, monitoring helps determine bottlenecks in the pre-processing and feature process pipelines, which could basically slow down the training process. The metric can be easily accessible using NVIDIA's command "nvidia-smi".

\subsection{GPU Memory Access and Utilization}
GPU memory is also one of the important metrics to understand how much GPU is being used in the process. The same command 'nvidia-smi' can provide a list of memory metrics that can be used for increasing the model training process. The GPU memory represents the percentage of time over the last second that the GPU's memory controller is being utilized to read or write from memory. These metrics help to tweak the batch size while training the models.

\subsection{Temperature and Power Usage}
Power consumption is also an important metric to determine the aspect of GPU performance. It indicates how power-intensive is the running application. Lower power is needed to deploy the machine learning models, especially on mobile devices. Therefore, it is significant to identify the power consumption. It is closely related to the ambient temperature the GPU is being used in. For any deep learning process, GPU's power consumption is important as thermal throttling at higher temperatures could slow the training process. It can also be monitored using commands from NVIDIA.

\subsection{Training}
The training time is the total time required for any deep learning model to forward and backward pass over the neural network architectures. The training is done by updating the weights and biases in the network. It is one of the metrics to benchmark the GPU's performance. GPU features like tuning the parameters and hyperparameters are extremely important in the training process.

\subsection{Throughput}
When the model is being deployed in production, the inference time is important. Inference time is similar to training time with the distinction that we only need forward passing of inputs and weights over the neural network. We will already have the trained model. The throughput is used to measure a GPU's performance in making the inferences faster. It is given by the number of samples processed per second by the model when GPU is being used. But, in a deep learning scenario, the metrics can be used depending on model architecture and deep learning applications. In CNN, the throughput can be calculated using images/second.

\subsection{Accuracy}
Accuracy is the most important metric while training and evaluating deep learning models. Two types of accuracy are monitored i.e. training accuracy and test accuracy. Although training accuracy doesn't play a significant role in determining how well the model performed, it sometimes helps in identifying whether the model overfits or underfits. When the model predicts all the actual positives as positives and actual negatives as negatives, then there will be no False Positive and False Negative cases. Then the model is said to be 100\% accurate.

\section{Experiments}
The experiment is set up to find out how much resources CPU and GPU consumed while training a deep learning model. A Convolutional Neural Network\cite{yanlecun} is being built using the Pytorch framework. A pre-trained model densenet121\cite{densenet} is used and we only changed the last layers of the network to train on our dataset. We are classifying cat and dog images. The dataset is downloaded from the Kaggle which is open source and free to use. The network is trained on both CPU and GPU and Pytorch Profiler is used to visualize the performance in tensorboard. Along with these frameworks, NVIDIA commands are used to evaluate the GPU performance. CPU evaluation is done using the built-in operating system monitoring system. Following are the experimental settings that are setup to perform this experiment specifically for hardware.

\begin{itemize}
\item Operating System: Windows 11 x64
\item Manufacturer: Dell
\item CPU: Intel(R) Xeon(R) Gold 6256
\item GPU: NVIDIA A6000 (2)
\item Total GPU Memory: 48 GB (2)
\item Clock Rate: 3.60 GHZ
\item Total RAM: 512 GB
\item Total Disk: 8 TB SSD
\end{itemize}

Regarding the neural network architectures and the dataset used, we downloaded the dataset from Kaggle. There are a total of 8005 training sets and 2023 test sets of cat and dog images. A trained CNN network 'densenet121' is used to define the architecture. Transfer learning is used to train the network as it helps in reducing the training time so that we can focus more time on evaluation. Different hyperparameters are tweaked to find the response of CPU and GPU resource utilization. Following are the settings for the convolutional neural network model.

\begin{figure}[hbt!]
  \centering
  \includegraphics[scale=0.35]{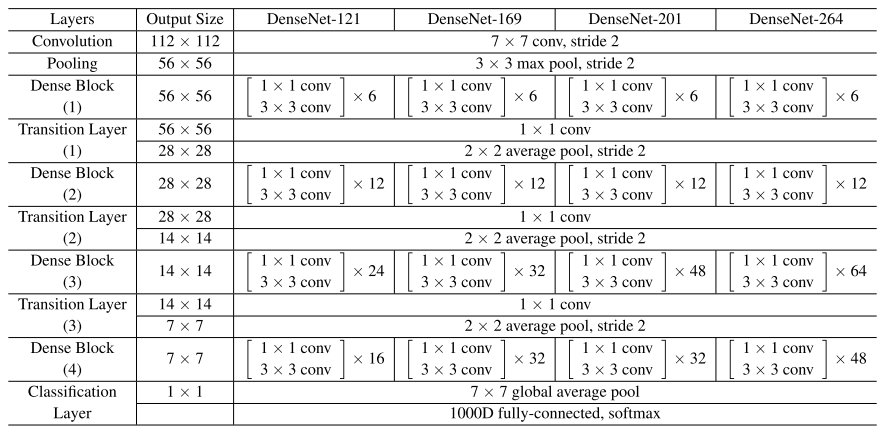}
  \caption{DenseNet121 architecture\cite{densenet}}
  \end{figure}

\begin{itemize}
\item Model: DenseNet121 \cite{densenet}
\item Total Training set: 8005
\item Total Testing set: 2023
\item Total classes: 2
\item Total epochs: 20
\item Optimizer: Adam
\item Loss Function: Cross-entropy
\item Batch size: 64 and 128
\item Learning Rate: 0.003 and 0.03
\end{itemize}

\section{Discussions and Results}
\vspace{0.5cm}
The CNN model is trained for 20 epochs with some hyperparameters tweaking. In CPU and GPU, the metrics are evaluated carefully.

\subsection{Utilization}
With a batch size of 64 and a learning rate of 0.003, the tasks are distributed over two GPU. Only 3\% of the GPU:0 and 14\% of the second GPU:1 was consumed for training the model as shown in Figure 4. It shows the minimum utilization of the GPU model. The task was first processed through CPU which uses ~62\% resource consumption. 

After increasing the batch size by 64 i.e. total batch size is 128, there is more utilization of GPU as compared to before. GPU:0 takes 69\% of its total memory whereas GPU:1 takes 14\% of its total memory as shown in Figure 7. This shows us that knowing the GPU utilization metrics helps us to change the hyperparameters inside the neural network that always help to optimize the network.

Regarding CPU, the total resource utilization is (60-63)\%. A total of 307 processes and 5410 threads were running with a clock speed of 4.10 GHz as shown in Figure 9. However, using only CPU, the model training time is comparatively higher as compared to GPU. It shows that for the training of deep neural networks, GPU utilization is needed in the maximum amount for faster and more efficient operations. From Figure 5, we see that most of the time CPU was spent on the training. There is no significant amount of CPU utilization while loading the data.

\begin{figure}[hbt!]
  \centering
  \includegraphics[scale=0.14]{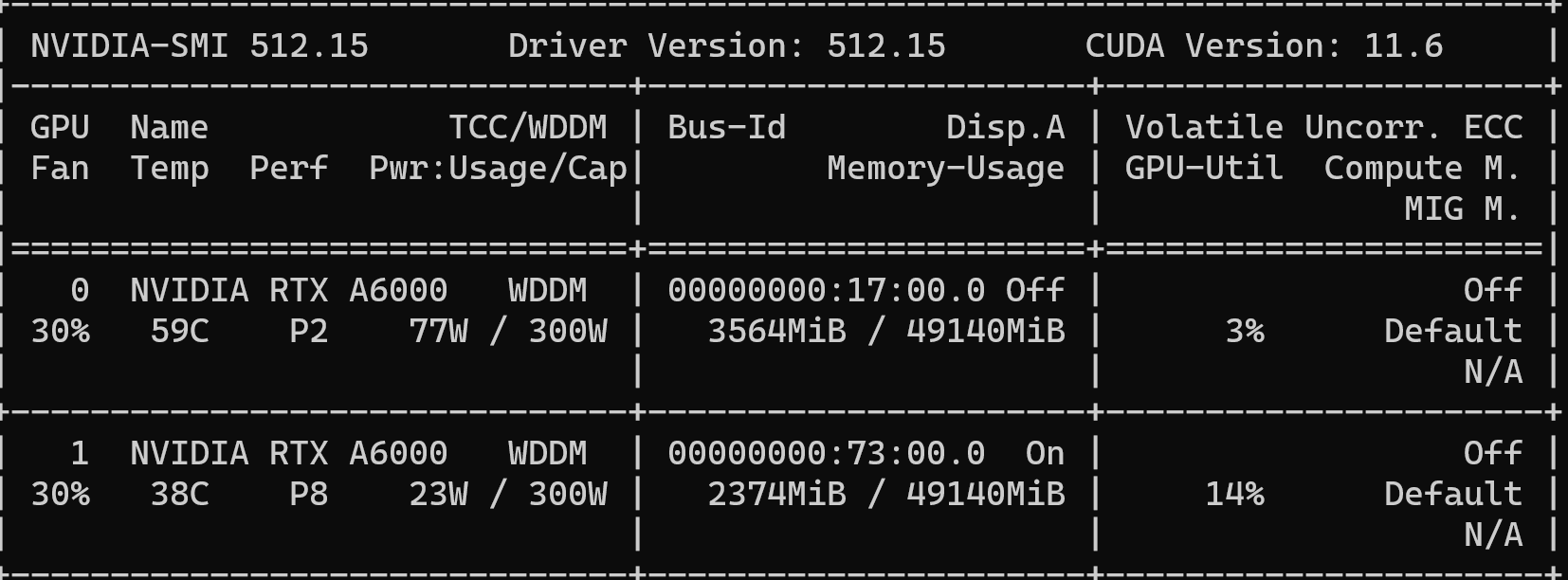}
  \caption{GPU Utilization for batch size 64}
  \end{figure}
  
\begin{figure}[hbt!]
\centering
\includegraphics[scale=0.24]{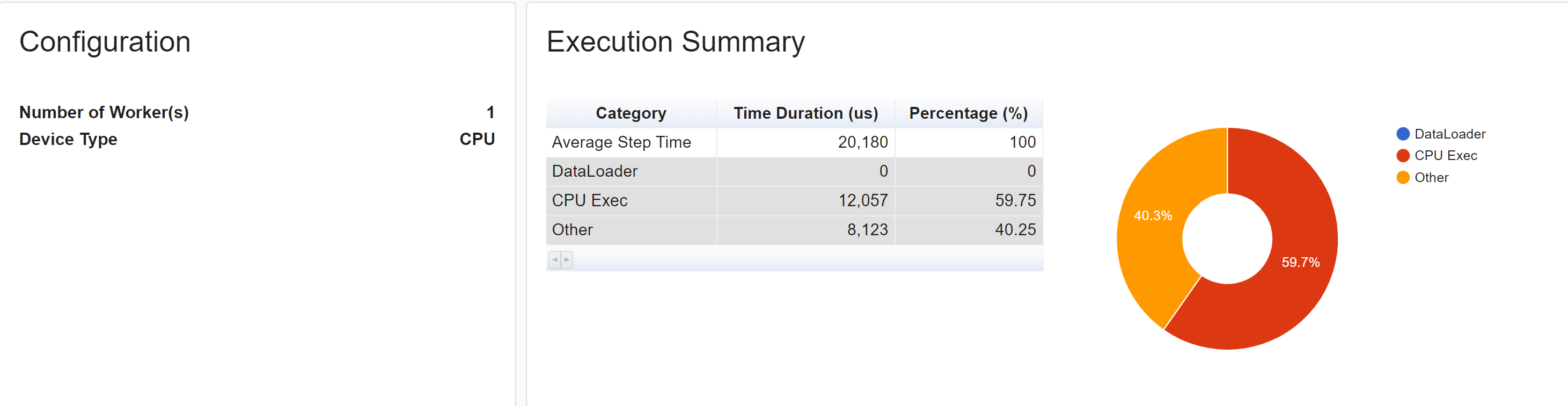}
\caption{CPU execution while training}
\end{figure}

\begin{figure}[hbt!]
\centering
\includegraphics[scale=0.24]{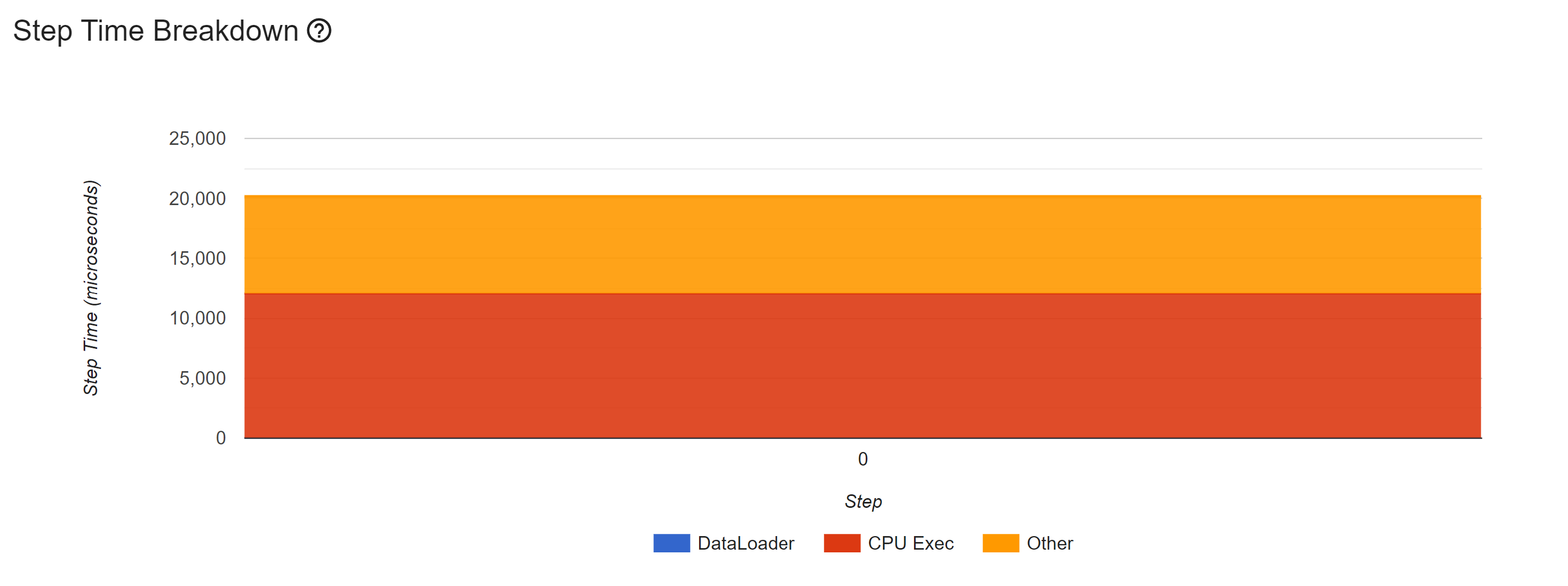}
\caption{Step Time Breakdown for CPU}
\end{figure}

\subsection{Memory Access and Utilization}
GPU shows a considerable amount of memory utilization when training on a higher batch size as compared to a lower batch size. However, the change in hyperparameter value doesn't have an effect on CPU memory utilization. When trained on batch size 64, the total memory consumption for GPU:0 is only around 3.4 GB out of 48 GB and GPU:1 is around 2.2 GB out of 48 GB. This shows a poor memory utilization of GPU when using a lower batch size.

After using a batch size of 128, GPU:0 utilizes around 5GB memory out of 48 GB. Similarly, GPU:1 utilizes around 1.7 GB out of 48 GB as shown in Figure 7. On the whole, the total memory consumption is greater as compared to the lower batch size. This shows that we can optimize the memory inside the GPU after profiling its memory consumption.

For CPU, there is no effect in memory consumption with hyperparameters value. The total amount of memory consumption is 32 GB out of 512 GB when trained on CPU only as shown in Figure 8.

\begin{figure}[hbt!]
  \centering
  \includegraphics[scale=0.40]{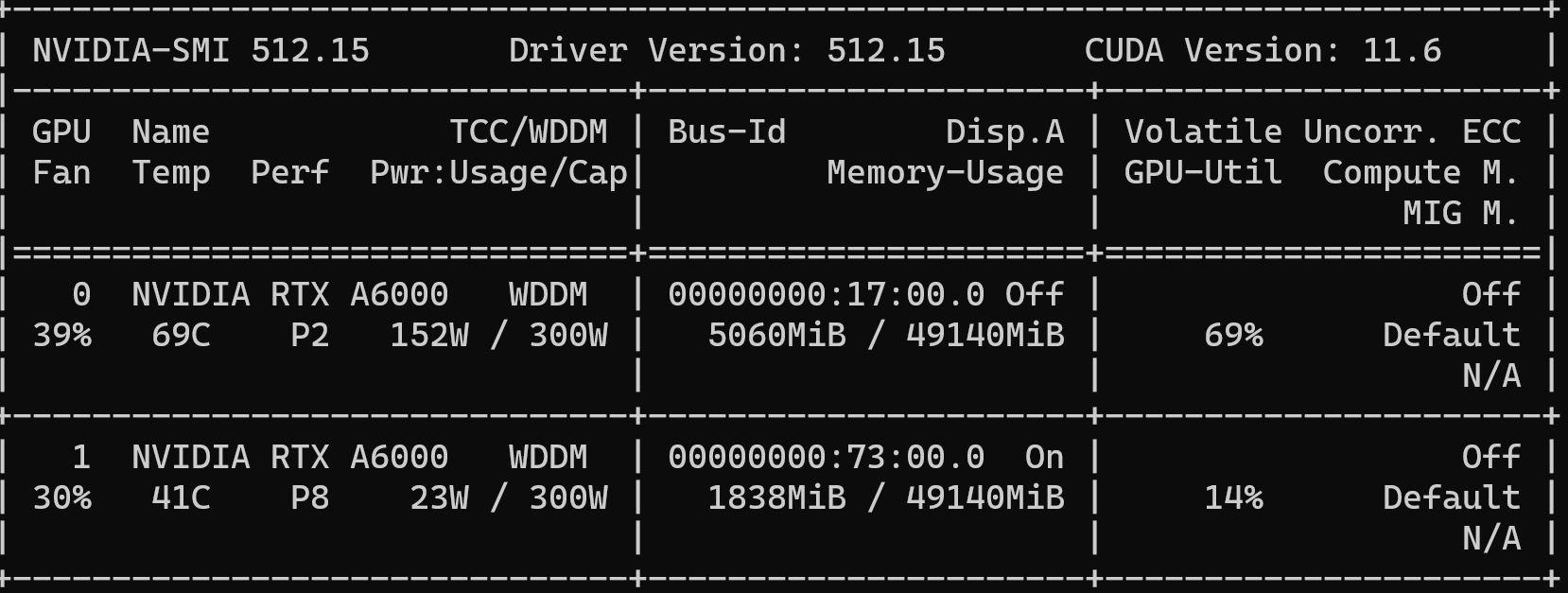}
  \caption{GPU utilization for batch size 128}
  \end{figure}
  
\begin{figure}[hbt!]
  \centering
  \includegraphics[scale=0.40]{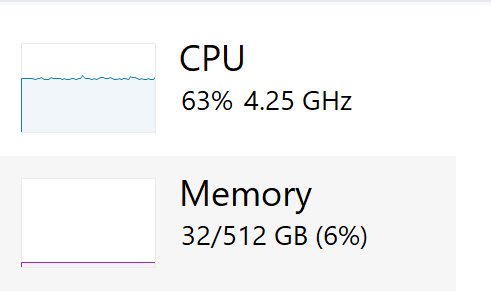}
  \caption{CPU memory usage}
  \end{figure}
  
\begin{figure}[hbt!]
  \centering
  \includegraphics[scale=0.10]{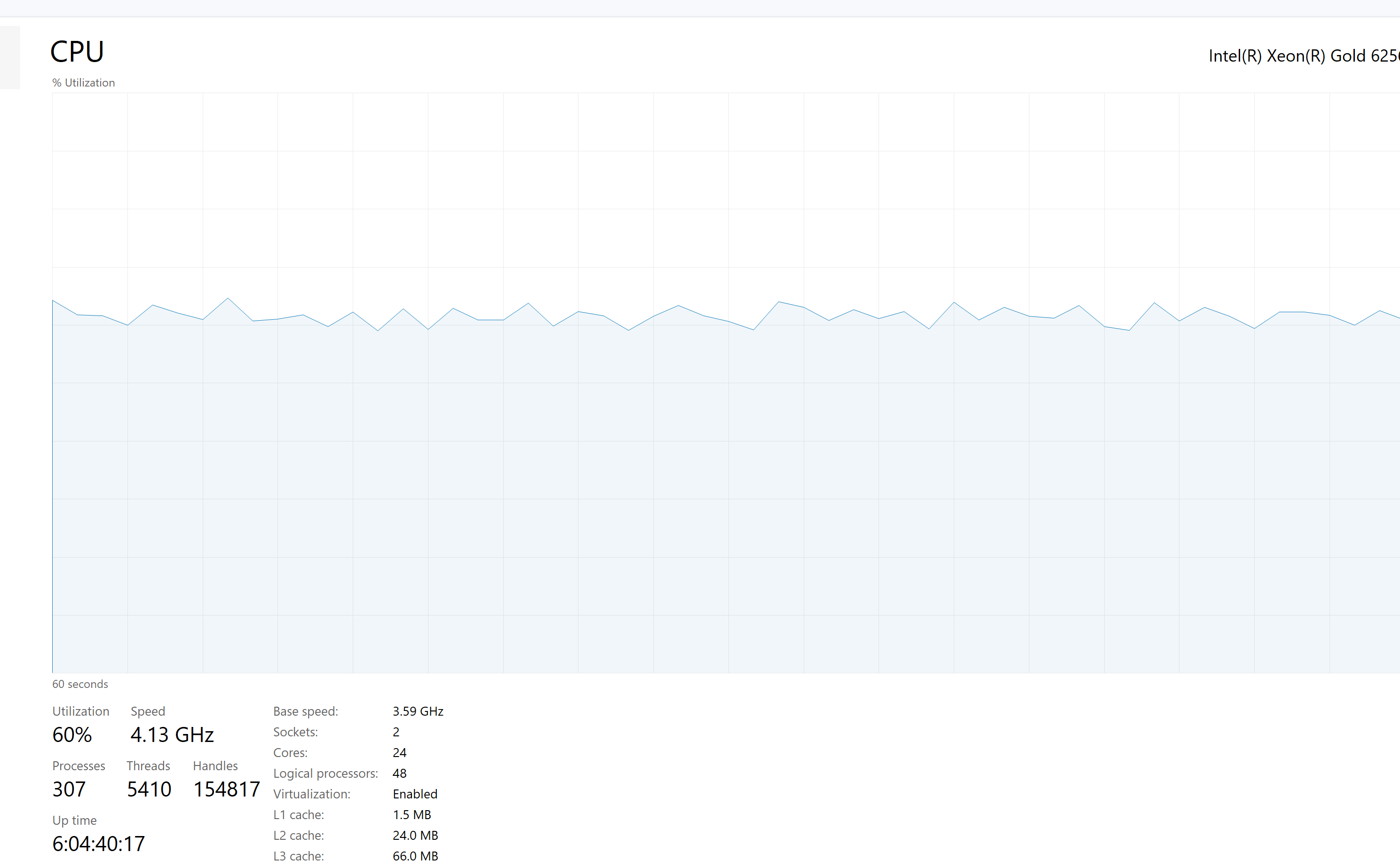}
  \caption{CPU Process and Thread}
  \end{figure}

\subsection{Temperature and Power Usage}
We also monitored the total temperature and Power usage of both CPU and GPU while training the model. The temperature for GPU 0 reaches up to 59 degrees Celsius and GPU 1 reaches up to 38 degrees Celsius for a batch size of 64 which is considered normal as shown in Fig 4.

Regarding batch size of 128, GPU:0 reaches the maximum temperature of 69 degrees Celsius and GPU:1 has a temperature of 41 degrees Celsius. Power consumption for GPU:0 is 152 W out of 300 W and only 23W out of 300 W for GPU:1 as shown in Fig 7.

Regarding CPU, the temperature ranges between 60-70 degrees Celsius regardless of the change in batch size as shown in Fig 9.

\begin{figure}[hbt!]
  \centering
  \includegraphics[scale=0.20]{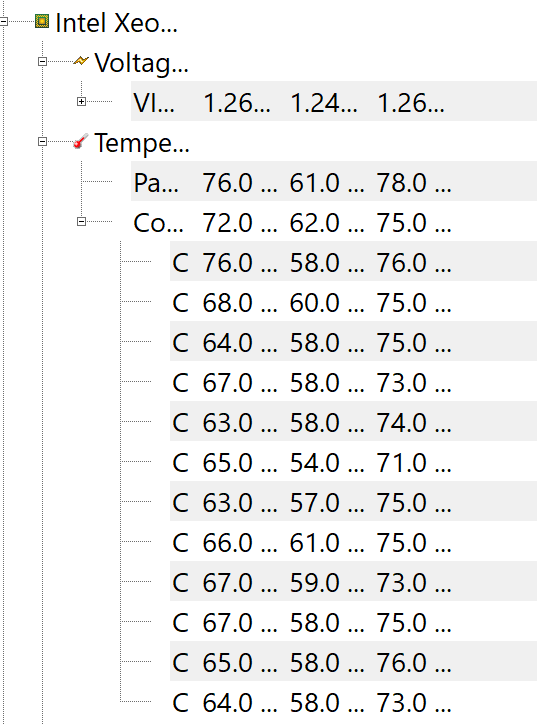}
  \caption{Power Utilization for CPU}
  \end{figure}

\subsection{Training Time}
The training time varies a lot as compared to GPU and CPU. We experimented with training the dog-cat classifier for 20 epochs for both CPU and GPU. The total training time for the CPU is around 13 hours for 20 epochs. However, it took only around 2 hours to train in GPU taking a batch size of 64. The test accuracy seems to be 99\% in both cases. This infers that the usage of GPU has a huge advantage while training the model. The most time consumed while training is to copy the data to be trained using CUDA as shown in Figure 11. The 'to' operator is used to copy the necessary data, loss function, and model to the GPU device. In addition, CNN layers also consume more time during execution which increases the training time.

When taking the batch size of 128, the training time reduced to around 1 hour 15 minutes which shows that GPU is consumed more to speed up the process without hampering the accuracy.

\begin{figure}[hbt!]
\centering
\includegraphics[scale=0.24]{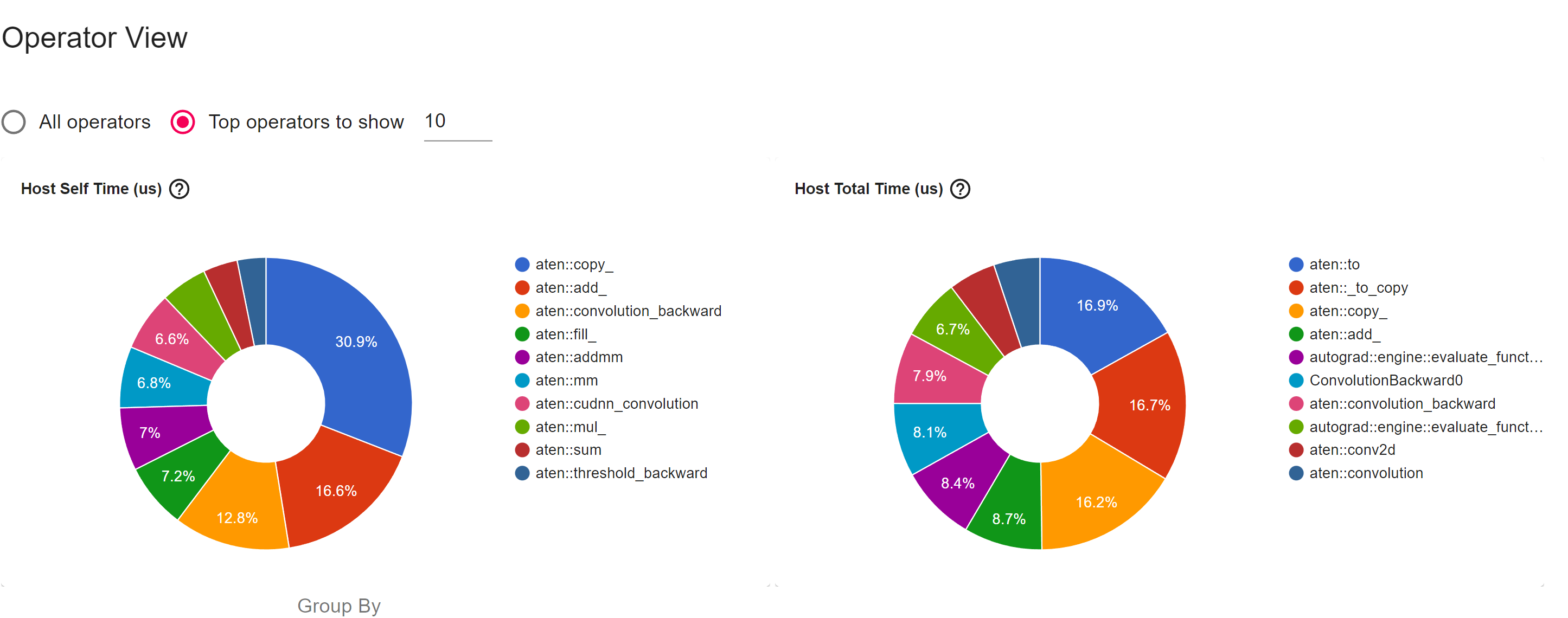}
\caption{Operator-wise execution breakdown}
\end{figure}

\subsection{Throughput}
The throughput is measured from the inference time. The inference time is greater in CPU as compared to GPU. The model was trained on both CPU and GPU and saved its weights for inference. In CPU, the testing time for one image is around 5 sec whereas in GPU it takes around 2-3 seconds which is better compared to CPU. This shows that GPU also plays a significant role in inference time that affects the throughput of the network.

\subsection{Accuracy}
The test accuracy seems to be similar on both CPU and GPU. Although it takes a lot of time to train the model on CPU, there are no significant differences in the test accuracy while testing on CPU and GPU. The test accuracy on both CPU and GPU is around (98-99)\%.

\section{Conclusion}
We have conducted experiments to show how the deep learning model CPU and GPU impact the time and memory consumption of CPU and GPU. A lot of factors impact artificial neural network training. We have studied the performance of different metrics regarding CPU and GPU usage. From the experiment, we also conclude that some parameters of the neural network model are also responsible for resource consumption across CPU and GPU. After profiling, we know how the performance is impacted by the use of CPU and GPU to handle the optimization procedure.

% \bibliographystyle{abbrvnat} % Choose a style. Others include 'unsrt', 'abbrv', etc.
% \bibliography{main}

\bibliographystyle{plain}
\bibliography{myreferences}

\end{document}